\documentclass{article}
\usepackage{spconf,amsmath,amssymb,graphicx}
\usepackage{mathtools}
\usepackage{microtype}
\usepackage{booktabs}
\usepackage[tight-spacing=true]{siunitx}
\usepackage{lipsum}
\usepackage{flushend}
\usepackage{subcaption}
\usepackage{multirow}
\usepackage{tabularray}
\UseTblrLibrary{booktabs}

\usepackage[backend=biber, bibencoding=utf8,
            style=ieee,
            maxbibnames=99,
            doi=false,
            url=false,
            isbn=false]{biblatex}
\AtEveryBibitem{%
  \ifentrytype{inproceedings}{%
    \clearfield{pages}%
    \clearlist{publisher}%
    \clearlist{organization}%
    \clearlist{location}}{}%
  \ifentrytype{misc}{%
    \clearfield{eprintclass}{} }}
    
\usepackage{tikzpagenodes} 

\usepackage[hidelinks]{hyperref}
\usepackage[capitalize]{cleveref}
\crefname{equation}{}{}
\usepackage{amssymb,amsmath}
\usepackage{mathtools}

\newcommand{\funcname}[1]{\mathrm{#1}}
\newcommand{\epow}[1]{\funcname{e}^{#1}}
\renewcommand{\jmath}{\mathrm{j}}

\newcommand{\numframes}{\ensuremath{L}}
\newcommand{\numbins}{\ensuremath{K}}

\newcommand{\mat}[1]{\mathbf{#1}}

\newcommand{\cleansig}{s}

\newcommand{\cleansigF}{\mat{S}}
\newcommand{\cleanmagF}{\mat{A}}

\DeclareMathOperator{\STFT}{STFT}
\DeclareMathOperator{\iSTFT}{iSTFT}

\newcommand{\eqcomma}{\,,} 
\newcommand{\eqperiod}{\,.} 
\newcommand{\minitimes}{{\mkern-2mu\times\mkern-2mu}}
\newcommand{\minirarrow}{{\mkern-2mu\rightarrow\mkern-2mu}}

\newcommand{\C}{\mathbb{C}}

\newcommand{\Cmat}[2]{\C^{#1\minitimes#2}}

\DeclarePairedDelimiter{\abs}{\lvert}{\rvert}
\newcommand{\phase}[1]{\mat{\Phi}_{#1}}

\newcommand{\estphase}[1]{\widehat{\mat{\Phi}}_{#1}}

\newcommand{\D}[1]{\mathrm{d}{#1}}
\newcommand{\submin}[0]{_\text{min}}
\newcommand{\submax}[0]{_\text{max}}

\DeclareMathOperator{\vecop}{vec}
\DeclarePairedDelimiter{\norm}{\lVert}{\rVert}

\title{DiffPhase: Generative Diffusion-based STFT Phase Retrieval}
\name{Tal Peer$^{1}$, Simon Welker$^{1,2}$, Timo Gerkmann$^{1}$\thanks{This work was funded by the Deutsche Forschungsgemeinschaft (DFG, German Research Foundation) --- project number 247465126, as well as by DASHH (Data Science in Hamburg - HELMHOLTZ Graduate School for the Structure of Matter) --- Grant- No. HIDSS-0002.}}
\address{$^{1}$ Signal Processing (SP), Universität Hamburg, Germany \\
      $^{2}$ Center for Free-Electron Laser Science, DESY, Hamburg, Germany}
\begin{document}
\ninept
\maketitle
\begin{abstract}
Diffusion probabilistic models have been recently used in a variety of tasks, including speech enhancement and synthesis. As a generative approach, diffusion models have been shown to be especially suitable for imputation problems, where missing data is generated based on existing data. Phase retrieval is inherently an imputation problem, where phase information has to be generated based on the given magnitude. In this work we build upon previous work in the speech domain, adapting a speech enhancement diffusion model specifically for STFT phase retrieval. Evaluation using speech quality and intelligibility metrics shows the diffusion approach is well-suited to the phase retrieval task, with performance surpassing both classical and modern methods.
\end{abstract}
\section{Introduction}\vspace{-2px}
\begin{tikzpicture}[remember picture,overlay]
  \node [draw=black, fill=white, text width=\textwidth, inner sep=3pt, outer sep=0, yshift=0cm] at (current page footer area){\footnotesize{Accepted paper. \copyright 2023 IEEE. Personal use of this material is permitted. Permission from IEEE must be obtained for all other uses, in any current or future media, including reprinting/republishing this material for advertising or promotional purposes, creating new collective works, for resale or redistribution to servers or lists, or reuse of any copyrighted component of this work in other works.}};
\end{tikzpicture}
\label{sec:intro}
Phase retrieval is a long-standing problem in the field of speech and audio processing. Regardless of the specific task (e.g. speech enhancement, source separation, speech synthesis), many systems do not operate directly on the time domain signal, but rather on the time-frequency representation provided by the short-time Fourier transform (STFT). Since the STFT of a real-valued signal is in general complex-valued, it can be separated into its magnitude and phase components, referred to here as \emph{magnitude spectrogram} and \emph{phase spectrogram} respectively. Although this view has been challenged in recent years, the magnitude spectrogram has been historically considered  more relevant and tangible, and therefore many STFT-based algorithms operate solely on it \cite{wangUnimportancePhaseSpeech1982,gerkmannPhaseProcessingSingleChannel2015}. However, the inverse transformation back to the time domain requires both components of the complex-valued STFT representation. Some tasks allow the use of some phase information (e.g. the noisy phase in the case of speech enhancement), while in other tasks, such as speech synthesis, the phase information might be completely missing. Even if some phase information is available, its combination with the estimated magnitude spectrogram is likely less than optimal. Phase retrieval methods aim to generate a phase spectrogram that matches a given magnitude spectrogram, either from scratch or based on some prior phase information.

Existing algorithms for STFT phase retrieval are plentiful and include ones based on iterative projections \cite{griffinSignalEstimationModified1984,masuyamaGriffinLimPhase2019a}, integration of phase derivatives \cite{prusaNoniterativeMethodReconstruction2017}, model-based approaches \cite{beauregardSinglePassSpectrogram2015} and more. The increasing popularity of deep neural networks (DNNs) has also affected the field of phase retrieval, and several DNN-based approaches have been proposed, e.g. involving phase derivatives \cite{takamichiPhaseReconstructionAmplitude2018,masuyamaPhaseReconstructionBased2020}, an augmentation of the iterative projection method \cite{masuyamaDeepGriffinLim2021}, or a generative approach using a generative adversarial network (GAN) \cite{oyamadaGenerativeAdversarialNetworkbased2018a}.

Diffusion-based generative models are a recent innovation that has already been shown to be very effective on various computer vision tasks, including inpainting \cite{lugmayrRePaintInpaintingUsing2022}, super-resolution \cite{sahariaImageSuperResolutionIterative2022}, text-to-image mapping \cite{sahariaPhotorealisticTexttoImageDiffusion2022} and more \cite{yangDiffusionModelsComprehensive2022}. Their application is, however, not limited to vision tasks and they have already been used in other fields, including text-to-speech \cite{popovGradTTSDiffusionProbabilistic2021}, as well as speech enhancement and dereverberation \cite{luStudySpeechEnhancement2021a, luConditionalDiffusionProbabilistic2022, welkerSpeechEnhancementScoreBased2022, richterSpeechEnhancementDereverberation2022}. The basic idea is gradual addition of noise to a clean sample until a completely corrupted sample is obtained. A DNN is then trained to invert the noise-addition process and thus implicitly learns the clean data distribution. The trained DNN can then be used for either unconditional generation, i.e. converting a noise sample to a sample corresponding to the clean data distribution, or for conditional generation which uses an auxiliary input that restricts the resulting distribution. The latter is more relevant here since we aim to generate a phase spectrogram conditioned on a given magnitude spectrogram. 

While several variants of the diffusion framework exist, we focus here on the formulation based on stochastic differential equations (SDEs) as proposed by Song et al. \cite{songScorebasedGenerativeModeling2021}. This framework has been adapted and extended to the conditional task of speech enhancement in the complex STFT domain \cite{welkerSpeechEnhancementScoreBased2022,richterSpeechEnhancementDereverberation2022} with very promising results. Since phase retrieval can be considered an imputation task (i.e., reconstruction of missing data), it seems only natural to approach it using a generative method such as diffusion models. In this work, we propose to modify an existing speech enhancement diffusion model and apply it to the STFT phase retrieval task by considering the loss of phase information (instead of the addition of environmental noise) as an auxiliary corruption process. Considering speech signals in particular, we show that the resulting phase retrieval method (``DiffPhase'') is able to surpass classical algorithms as well as DNN-based methods in terms of perceptual metrics.
\begin{figure*}[ht]
    \centering
    \begin{subfigure}{.35\textwidth}
        \includegraphics[width=0.99\textwidth]{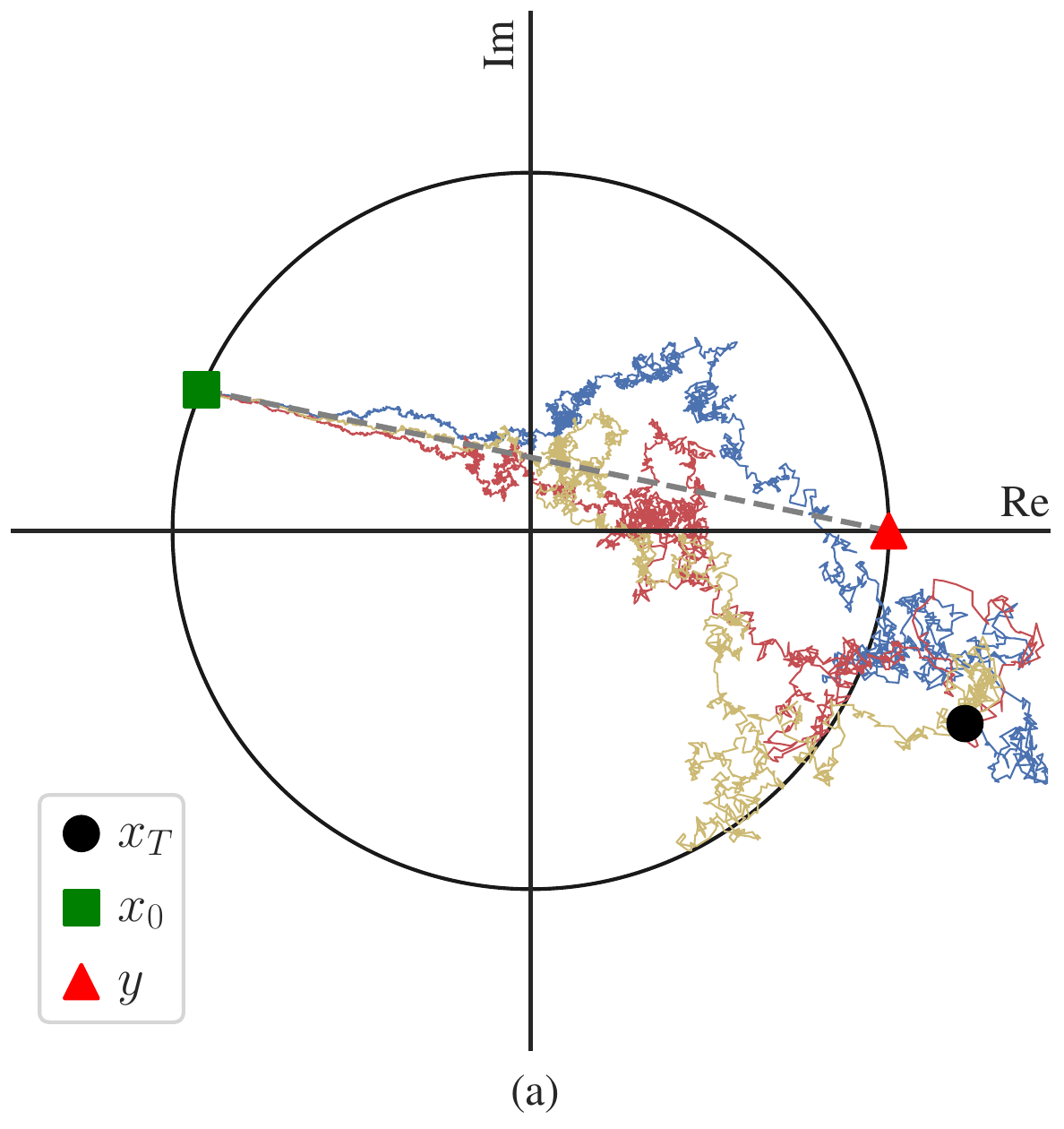}
        \phantomcaption\label{fig:traj}
    \end{subfigure}\hspace{1cm}
    \begin{subfigure}{.4265\textwidth}
        \includegraphics[width=0.99\textwidth]{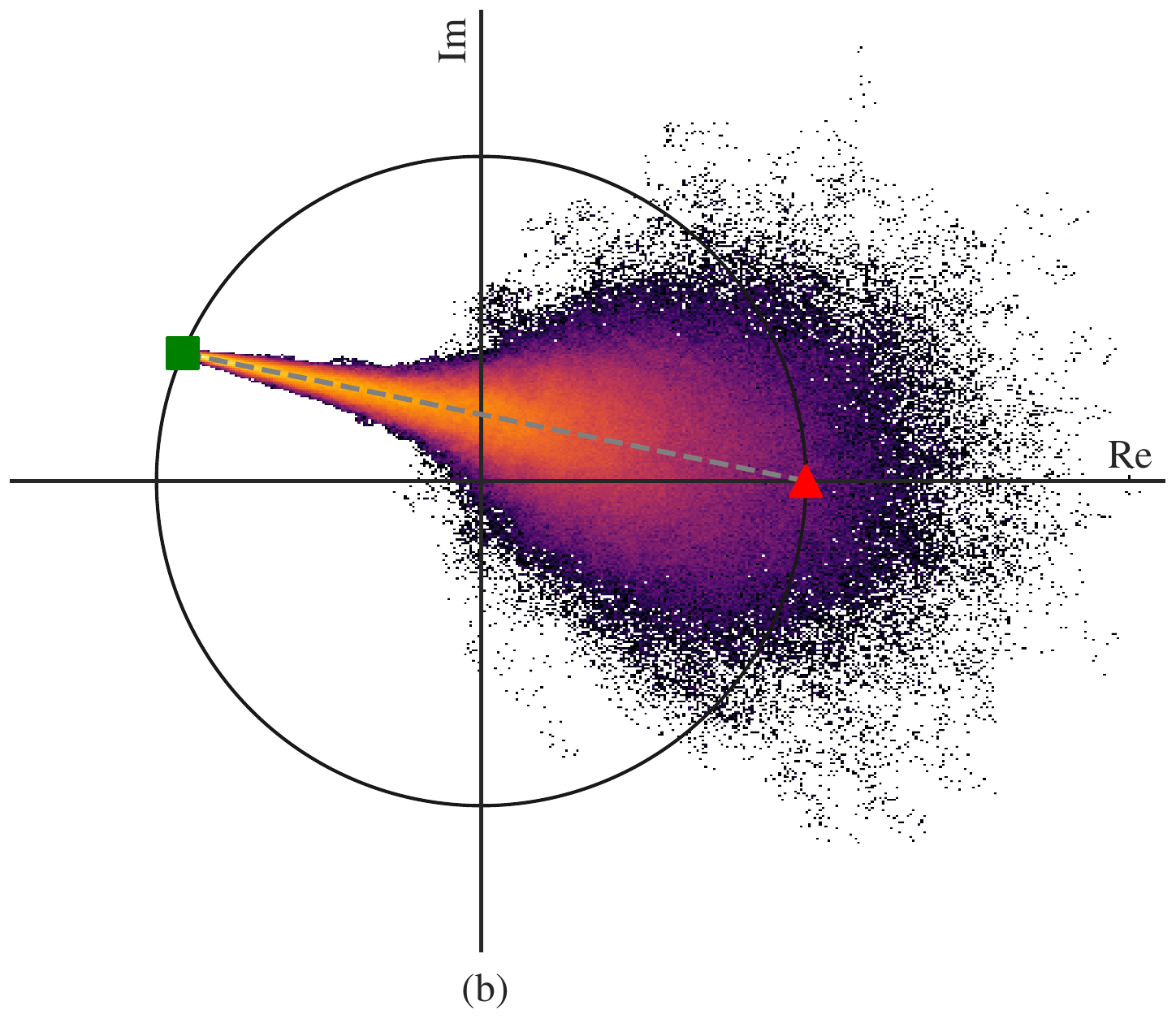}
        \phantomcaption\label{fig:hist}
    \end{subfigure}\vspace{-10px}
    \caption{Illustration of the reverse process, as given in \cref{eq:sde_reverse}, but with scalar $x_t, y \in \mathbb{C}$ and assuming the score is perfectly estimated. Note that at each time-step, the mean $\mu(x_0, y, t)$ lies on the dashed line (see \cref{eq:sde_mean}). \subref{fig:traj} Three different simulated trajectories of the reverse process starting from the same final state $x_T$. \subref{fig:hist} A histogram of all process states $x_t$ in 1000 different realizations of the reverse process. Brighter color corresponds to higher histogram values.}
    \label{fig:traj_hist}
\end{figure*}
\section{STFT phase retrieval}\vspace{-2px}
\label{sec:pr}
We denote the STFT (complex spectrogram) of a time domain signal $\cleansig$ as $\cleansigF \in \Cmat{\numbins}{\numframes}$, which is represented in polar coordinates by the magnitude spectrogram $\cleanmagF = \abs{\cleansigF}$ and phase spectrogram $\phase{\cleansigF}$:
\begin{equation}
    \label{eq:stft_basic}
    \cleansigF = \cleanmagF \epow{j \phase{\cleansigF}} \eqperiod
\end{equation}
Note that we always refer to the entire spectrogram and thus refrain from explicitly using indices to denote time frames and frequency bins. $\numbins$ and $\numframes$ denote the number of frequency bins and time frames, respectively.

Phase retrieval algorithms are tasked with finding an estimated phase spectrogram $\estphase{\cleansigF}$, given a known magnitude spectrogram $\cleanmagF$ and possibly some prior phase information $\estphase{\cleansigF}^0$. Due to overlapping frames, the STFT representation is inherently redundant. This redundancy gives rise to the notion of \emph{consistency}, where a complex spectrogram $\cleansigF$ is said to be consistent if $\cleansigF = \STFT(\iSTFT(\cleansigF))$ and only consistent spectrograms correspond to a time-domain signal \cite{lerouxConsistentWienerFiltering2013a,gerkmannPhaseProcessingSingleChannel2015}. Formulated as a constraint, consistency forms the foundation for a class of phase retrieval algorithms based on \emph{iterative projections}. The simplest of these is the Griffin-Lim algorithm (GLA) \cite{griffinSignalEstimationModified1984}, in which two constraints are enforced in each iteration: the known magnitude spectrogram and the consistency constraint. Several extensions of GLA exist (e.g., \cite{perraudinFastGriffinLimAlgorithm2013,peerGriffinLimImprovedIterative2022a}) but they are all based on the same principle of iterative projection. Other approaches to STFT phase retrieval do not explicitly consider consistency, but still rely on the above redundancy. Data-driven approaches, such as \cite{masuyamaDeepGriffinLim2021,masuyamaPhaseReconstructionBased2020}, as well as this work, use machine learning algorithms in order to implicitly learn a mapping from a phase-less magnitude spectrogram to a complex spectrogram. 

While some prior phase information might be available, in the most general case the phase is entirely unknown. In this case, it is possible to consider different initialization schemes for the different algorithms, e.g., uniformly sampled random phase. For simplicity and without loss of generality, we assume a zero-phase initialization throughout this paper.
\section{Diffusion-based generative models}\vspace{-2px}
\label{sec:diffusion}
A DNN-based diffusion model consists of a \emph{forward process} which adds noise to a sample, a \emph{reverse process} which removes noise, and a DNN which estimates the noise to be removed at each step of the reverse process. 
\subsection{Forward process}\vspace{-2px}
We consider a stochastic forward process defined by an SDE of the general form
\begin{equation}
    \label{eq:sde_forward}
        \D{\mathbf x_t} = \mathbf{f}(\mathbf{x}_t, \mathbf{y}, t) \D{t} + g(t) \D{\mathbf{w}} \eqcomma 
\end{equation}
where $\mathbf{w}$ is the standard complex Wiener process, $\mathbf{x}_t$ is the current process state, $t \in [0,T]$ is a continuous time-step variable expressing the current progress of the process (note that $t$ is completely unrelated to the time dimension of any signal in the time or time-frequency domains), and $\mathbf{f}(\mathbf{x}_t, \mathbf{y}, t)$, $g(t)$ are the \emph{drift} and \emph{diffusion} coefficients, respectively. 

The forward process turns a clean sample $\mathbf{x}_0$ into a completely corrupted sample $\mathbf{x}_T$ by gradually adding noise, composed of Gaussian noise from the Wiener process as well as auxiliary corruption given by the drift term. Following \cite{welkerSpeechEnhancementScoreBased2022,richterSpeechEnhancementDereverberation2022}, we use the Ornstein-Uhlenbeck Variance Exploding (OUVE) SDE and define the drift and diffusion coefficients accordingly:
\begin{equation}
    \label{eq:drift_coeff}
    \mathbf{f}(\mathbf{x}_t, \mathbf{y}, t) = \gamma(\mathbf y-\mathbf x_t) \eqcomma
\end{equation}
\begin{equation}
    \label{eq:diff_coeff}
    g(t) = \left[ \sigma\submin \left(\frac{\sigma\submax}{\sigma\submin}\right)^t \sqrt{2\log\left(\frac{\sigma\submax}{\sigma\submin}\right)} \right] \eqcomma
\end{equation}
where $\sigma\submin$, $\sigma\submax$ and $\gamma$ are constant scalar parameters.

In the context of phase retrieval, the clean sample corresponds to the full complex spectrogram with known magnitude and phase. The auxiliary corruption process represents gradual loss of phase information. We thus define 
\begin{align}
    \mathbf{x}_0 &:= \vecop(\cleansigF) \eqcomma\\
    \mathbf{y} &:= \vecop(\cleanmagF) \eqcomma
\end{align}
where $\vecop(\cdot)$ converts a matrix to a vector by stacking the columns. 
\subsection{Reverse process}\vspace{-2px}
The forward process ($0\, \minirarrow\, T$) described in \cref{eq:sde_forward}, has an associated reverse process ($T\, \minirarrow\, 0$), given by the SDE \cite{andersonReversetimeDiffusionEquation1982}
\begin{equation}
    \label{eq:sde_reverse}
        \D{\mathbf x_t} =
        \left[
            -\mathbf f(\mathbf x_t, \mathbf y, t) + g(t)^2\nabla_{\mathbf x_t} \log p_t(\mathbf x_t|\mathbf y)
        \right] \D{t}
        + g(t)\D{\bar{\mathbf w}} \eqcomma
\end{equation}
where $\bar{\mathbf{w}}$ denotes the time-reversed Wiener process. Considering \cref{eq:drift_coeff,eq:diff_coeff}, the only unknown term present in \cref{eq:sde_reverse} is $\nabla_{\mathbf x_t} \log p_t(\mathbf x_t|\mathbf y)$ which is referred to as the \emph{score function}. Diffusion models (also ``score-based models'') use a DNN $\mathbf{s}_\theta (\mathbf{x}_t, \mathbf{y}, t)$ to estimate the score, therefore obtaining a tractable reverse SDE which can be solved by various numerical methods \cite{songScorebasedGenerativeModeling2021} to produce an estimate $\widehat{\mathbf{x}}_0$ of the initial state $\mathbf{x}_0$. 

An illustration of the reverse process for a simplified scalar problem is presented in \cref{fig:traj_hist}, showing how the process converges in distribution towards the initial state.
\subsection{Training}\vspace{-2px}
\label{sec:training}
The score-estimation DNN is trained using a technique known as \emph{denoising score matching} \cite{vincentConnectionScoreMatching2011}. The mean and variance of the forward process have closed-form expressions \cite{richterSpeechEnhancementDereverberation2022}
\begin{equation}
    \label{eq:sde_mean}
    \boldsymbol\mu(\mathbf x_0,\mathbf y, t) = \mathrm e^{-\gamma t} \mathbf x_0 + (1-\mathrm e^{-\gamma t}) \mathbf y
    \eqcomma
\end{equation}
and
\begin{equation}
\label{eq:sde_var}
        \sigma(t)^2 = \frac{
        \sigma\submin^2\left(\left(\frac{\sigma\submax}{\sigma\submin}\right)^{2t} - \mathrm e^{-2\gamma t}\right)\log(\frac{\sigma\submax}{\sigma\submin})
    }{\gamma+\log(\frac{\sigma\submax}{\sigma\submin})}
    \eqcomma
\end{equation}
which, given a random $t \in [0,T]$, a data pair $(\mathbf{x}_0, \mathbf{y})$, and a standard complex Gaussian sample $\mathbf{z} \sim \mathcal{N}_{\mathbb C}(\mathbf{z}; \mathbf{0}, \mathbf{I})$, allow convenient sampling from the forward process with
\begin{equation}
\mathbf x_t =  \boldsymbol\mu(\mathbf x_0,\mathbf y, t) + \sigma(t) \mathbf{z} \eqperiod
\end{equation}
This sample is passed to the DNN $\mathbf{s}_\theta$, which is trained by minimizing the loss function
\begin{equation}
    \label{eq:objective}
        \mathcal{L} = \norm*{\mathbf s_\theta(\mathbf x_t, \mathbf y, t) + \frac{\mathbf z}{\sigma(t)}}_2^2
    \eqperiod
\end{equation}
An in-depth description of the training process and a full derivation of the loss function \cref{eq:objective} are available in \cite{richterSpeechEnhancementDereverberation2022}.
\subsection{Phase retrieval by sampling the reverse process}\vspace{-2px}
The trained DNN provides a score estimate which together with \cref{eq:sde_reverse} yields an approximate reverse process. An estimate of the initial state $\widehat{\mathbf{x}}_0$ (corresponding to an estimated complex spectrogram) is determined by solving the reverse SDE using one of various numerical SDE solvers. The final state (i.e., the initial state of the reverse process) is given by
\begin{equation}
    \mathbf{x}_T \sim \mathcal{N}_{\mathbb{C}}(\mathbf{x}_T; \mathbf{y}, \sigma(T)^2 \mathbf{I}) \eqperiod
\end{equation}
Recall that here $\mathbf{y}$ corresponds to the phase-less magnitude spectrogram. Thus, $\mathbf{x}_T$ is simply the magnitude spectrogram with a large amount of complex Gaussian noise added. While more sophisticated approaches exist, in this work we use the simple \emph{reverse diffusion sampler} described in \cite{songScorebasedGenerativeModeling2021} to solve the reverse SDE, based on our experiments showing sufficient performance.

Although the known magnitude spectrogram is included both as a DNN input as well as explicitly in the reverse SDE, the reverse process is not specifically constrained to retain the known magnitude in the final estimate $\widehat{\mathbf{x}}_0$. Therefore, as a final step before the inverse STFT, the known magnitude spectrogram is enforced again.
\section{Implementation Details}\vspace{-2px}
Based on its demonstrated performance for image tasks \cite{songScorebasedGenerativeModeling2021}, as well as for speech enhancement and dereverberation \cite{richterSpeechEnhancementDereverberation2022}, we employ the Noise Conditional Score Network (NCSN++) model \cite{songScorebasedGenerativeModeling2021}, with adaptations to complex spectrograms as described in \cite{richterSpeechEnhancementDereverberation2022}. This model is in essence a specialized U-Net encoder-decoder structure with additional attention modules between certain layers. The input to the network consists of $\mathbf{x}_t$ and $\mathbf{y}$. Additionally, the time-step $t$ is fed into each layer in the network in the form of learned Fourier embeddings \cite{vaswaniAttentionAllYou2017}. The model is trained as described in \cref{sec:training}, by sampling a uniformly random $t$ and constructing $\mathbf{x}_t$ accordingly.

To facilitate the training process, model inputs are transformed using the approach proposed in \cite{welkerSpeechEnhancementScoreBased2022}, consisting of a fixed magnitude compression and scaling applied to each time-frequency bin:
\begin{equation}
    \label{eq:comp_transform}
    \widetilde{v} = \beta \abs{v}^\alpha \epow{j \phi_{v}} \eqcomma
\end{equation}
where $v$ represents a single bin and $\phi_v$ is its phase. Note that this transformation leaves the phase spectrogram unaltered.

To avoid numerical instability due to the numerator of \cref{eq:sde_var} becoming zero, we limit $t$ to the interval $[t_\varepsilon, T]$ during training and reverse sampling. The reverse SDE is solved using a simple discretization technique similar to the Euler-Maruyama method \cite[Appendix E]{songScorebasedGenerativeModeling2021}. This method uses a uniform discretization with step-size $\Delta t = T/N$. The choice of $N$ is non-trivial and has a large impact on the performance of the diffusion model, as shown in \cref{sec:results}.
\begin{table}[t]
    \centering
    \scalebox{0.95}{
   \begin{tblr}{width=\columnwidth, colspec={lX[c,m]X[c,m]}}
    \toprule
                                    &     DiffPhase & DiffPhase-small        \\
        \midrule
        Encoder resolutions         &     {$\{256,128,\dotsc,$ \\ $ 8,4\}$}   &     {$\{256,128,64,$ \\ $32,16\}$}     \\
         \# residual blocks           &     2    &      1     \\
        Attention  modules          &     At $16\minitimes 16$ and $4\minitimes4$    &      Only at $16 \minitimes 16$    \\
        \# parameters               &     \num{6.5e7} & \num{2.2e7} \\
        \bottomrule
    \end{tblr} }
    \caption{Configuration details of the NCSN++ architecture \cite{songScorebasedGenerativeModeling2021}. Encoder resolutions represent the output size of 2D convolutional layers in the U-Net encoder (the output is always square, i.e., a resolution of 256 means size 256$\minitimes$256). The decoder is built symmetrically and thus not listed. Each resolution consists of one or more residual blocks. The number of channels at each resolution is always 128 for the first two resolutions and 256 afterwards. Attention modules are present at the encoder-decoder bottleneck (i.e., after the smallest resolution) and optionally at a larger resolution too.}
    \label{tab:ncsn}
\end{table}
\begin{figure}[t]
    \centering
    \begin{subfigure}{.92\columnwidth}
        \includegraphics[width=\textwidth]{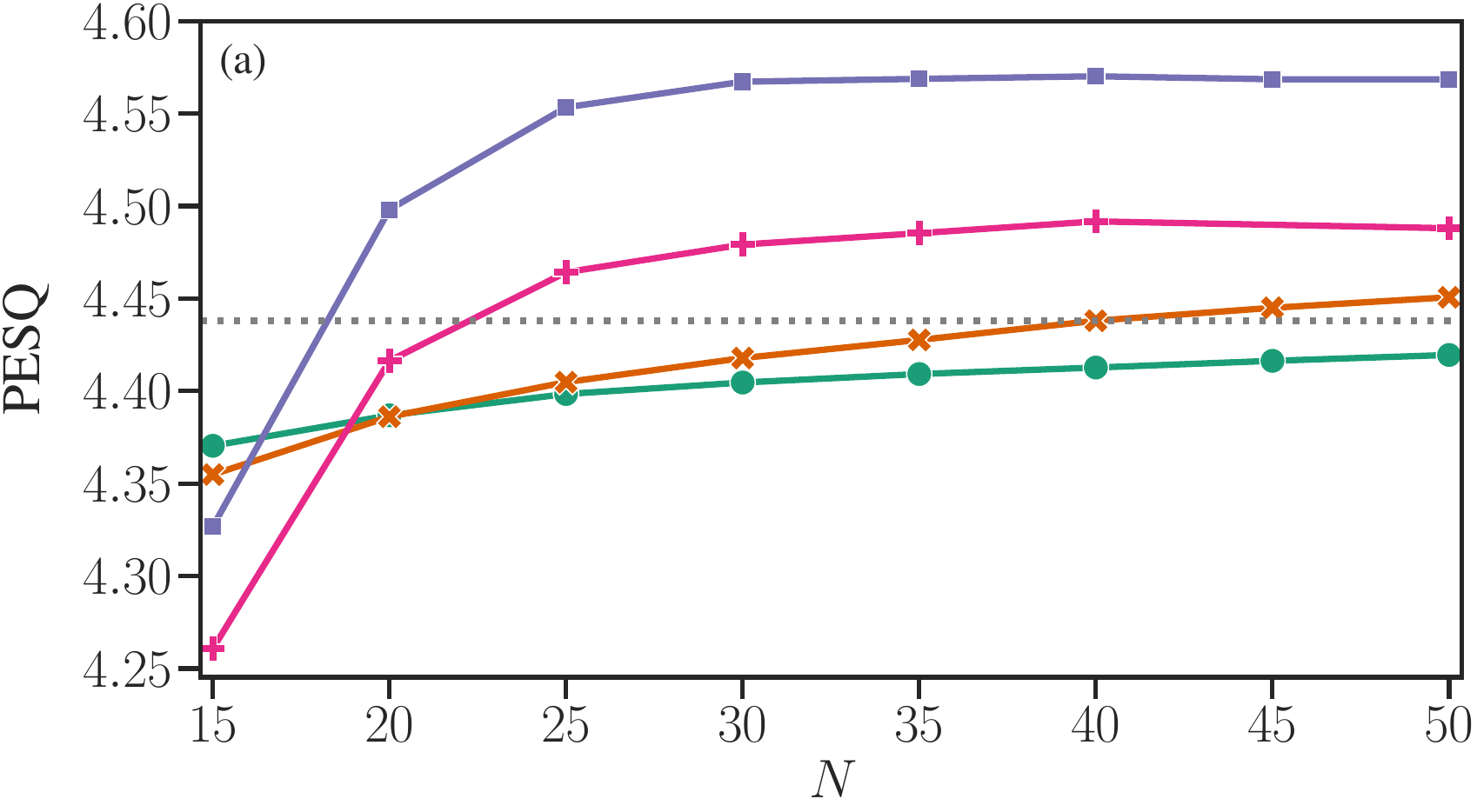}
        \phantomcaption\label{fig:pesq}
    \end{subfigure}\vspace{-2px}
    \begin{subfigure}{.92\columnwidth}
        \includegraphics[width=\textwidth]{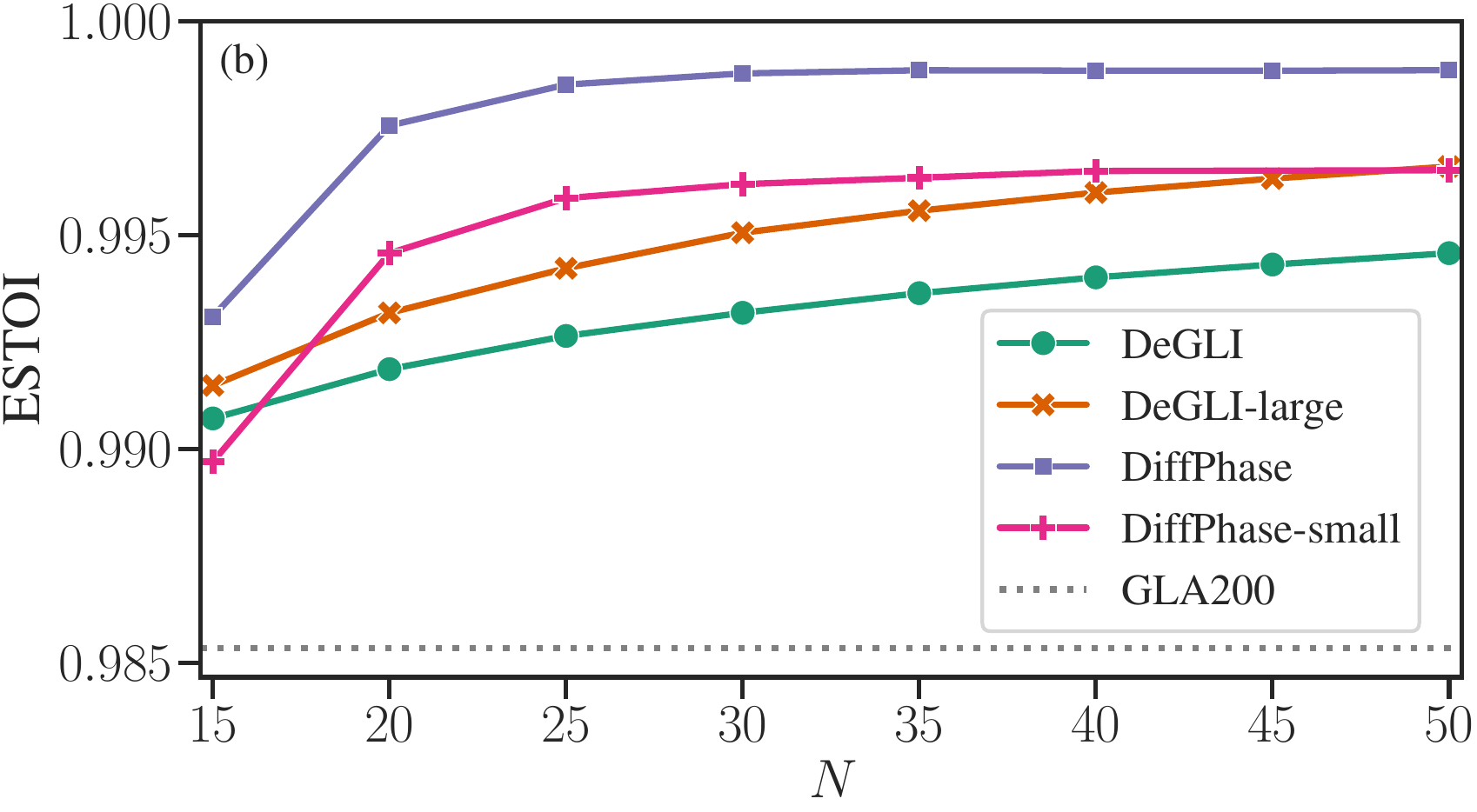}
        \phantomcaption\label{fig:estoi}
    \end{subfigure}\vspace{-9px}
    \caption{\subref{fig:pesq} and \subref{fig:estoi} Mean PESQ and ESTOI scores of clean speech reconstructed by different algorithms on the test set for different values of $N$. For DiffPhase models, $N$ is the number of discrete steps in the reverse process. For DeGLI, $N$ is the number of consecutive DeGLI blocks applied to the input. In both cases, $N$ is equal to the number of DNN evaluations. For comparison, we also include the scores achieved by GLA after 200 iterations as a dotted line.}
    \label{fig:pesq_estoi}
\end{figure}
\section{Experimental Evaluation}\vspace{-4px}
\label{sec:exp}
We evaluate the performance of the proposed diffusion model in terms of perceptual evaluation metrics (PESQ and ESTOI). We also evaluate the model in terms of runtime, i.e. the time it takes to retrieve the phase for an utterance of a given length.

Training and evaluation are performed using clean speech data, where the magnitude spectrogram is perfectly known. The phase spectrogram is completely missing and replaced by zeros. PESQ and ESTOI are calculated with the clean speech signal as reference. For runtime measurement, we used an NVIDIA GeForce RTX 2080 Ti GPU, with results averaged over 100 runs for each algorithm.
\subsection{Baselines}\vspace{-5px}
We compare the proposed approach (DiffPhase) to the classical Griffin-Lim algorithm (see \cref{sec:pr})---a non-learned method, and also to its DNN-based extension DeGLI \cite{masuyamaDeepGriffinLim2021}. While DiffPhase and DeGLI are both based on DNNs, they differ greatly in size. The NCSN++ network that we employ has about \num{6.5e7} parameters while DeGLI only about \num{4e5}. For this reason, we also consider two additional variants in our evaluation: DiffPhase-small with about \num{2e7} parameters and DeGLI-large with about \num{2e6}. Details of the different NCSN++ configurations for DiffPhase and DiffPhase-small are given in \cref{tab:ncsn}.

The DeGLI-large variant contains six inner layers instead of two and uses a kernel size of $5 \minitimes 5$ instead of $5 \minitimes 3$ (see \cite{masuyamaDeepGriffinLim2021} for details). We have also added residual connections between the inner layers to facilitate training. Note that extending DeGLI beyond this point with even more parameters did not result in improved performance.
\subsection{Data}\vspace{-5px}
For training of both DiffPhase and DeGLI we use clean speech data from a subset of the Voice Bank corpus \cite{veauxVoiceBankCorpus2013} composed of 28 speakers (14 male, 14 female) with about 400 utterances each, as described in \cite{valentini-botinhaoInvestigatingRNNbasedSpeech2016}. All speech data was downsampled to \qty{16}{\kilo\hertz}. Each utterance was transformed into a complex spectrogram with the STFT, using the same parameters as in \cite{richterSpeechEnhancementDereverberation2022} (a 510-sample Hann window with a frame shift of 128 samples). For each training step, a random spectrogram slice containing 256 frames was selected, resulting in an input spectrogram of size $256\minitimes 256$, which was subsequently transformed using \cref{eq:comp_transform} and input into the network as described in \cref{sec:training}. The DeGLI model was trained using the same data and STFT parameters, however without magnitude compression.

Evaluation was carried out on clean speech data from the WSJ0 corpus \cite{paulDesignWallStreet1992}, using the \texttt{si\_et\_05} test set, consisting of eight speakers and 651 utterances in total.
\subsection{Hyperparameters}\vspace{-5px}
The forward and reverse SDEs in \cref{eq:sde_forward,eq:sde_reverse} include several tunable parameters.
The hyperparameter values in this work are either based on other literature (mainly \cite{richterSpeechEnhancementDereverberation2022}) or on experimentation. The following values were used for SDE hyperparameters:
\begin{equation*}
    \sigma\submin = 0.05, \quad \sigma\submax = 0.5, \quad \gamma =1.5, \quad t_\varepsilon = 0.03, \quad T=1
\end{equation*}
Magnitude scaling and compression in \cref{eq:comp_transform} were applied with $\alpha=0.5,\beta=0.15$ 
\section{Results}
\label{sec:results}
Evaluation results in terms of mean PESQ and ESTOI scores are depicted in \cref{fig:pesq_estoi} for different values of $N$, where $N$ is the number of consecutive blocks in the case of DeGLI or the number of discrete steps in the reverse process for DiffPhase. For small $N$, DeGLI is superior but for a sufficiently large $N$ (e.g. 20), DiffPhase surpasses DeGLI and almost reaches a perceptually perfect reconstruction of the clean speech signal, showing that DiffPhase is in principle capable of high-quality phase retrieval, but only under sufficiently fine discretization (note that this might be a limitation of the simple SDE solver we have used here, and not necessarily an inherent drawback of the diffusion approach).

\begin{figure}[t]
    \centering
    \includegraphics[width=.92\columnwidth]{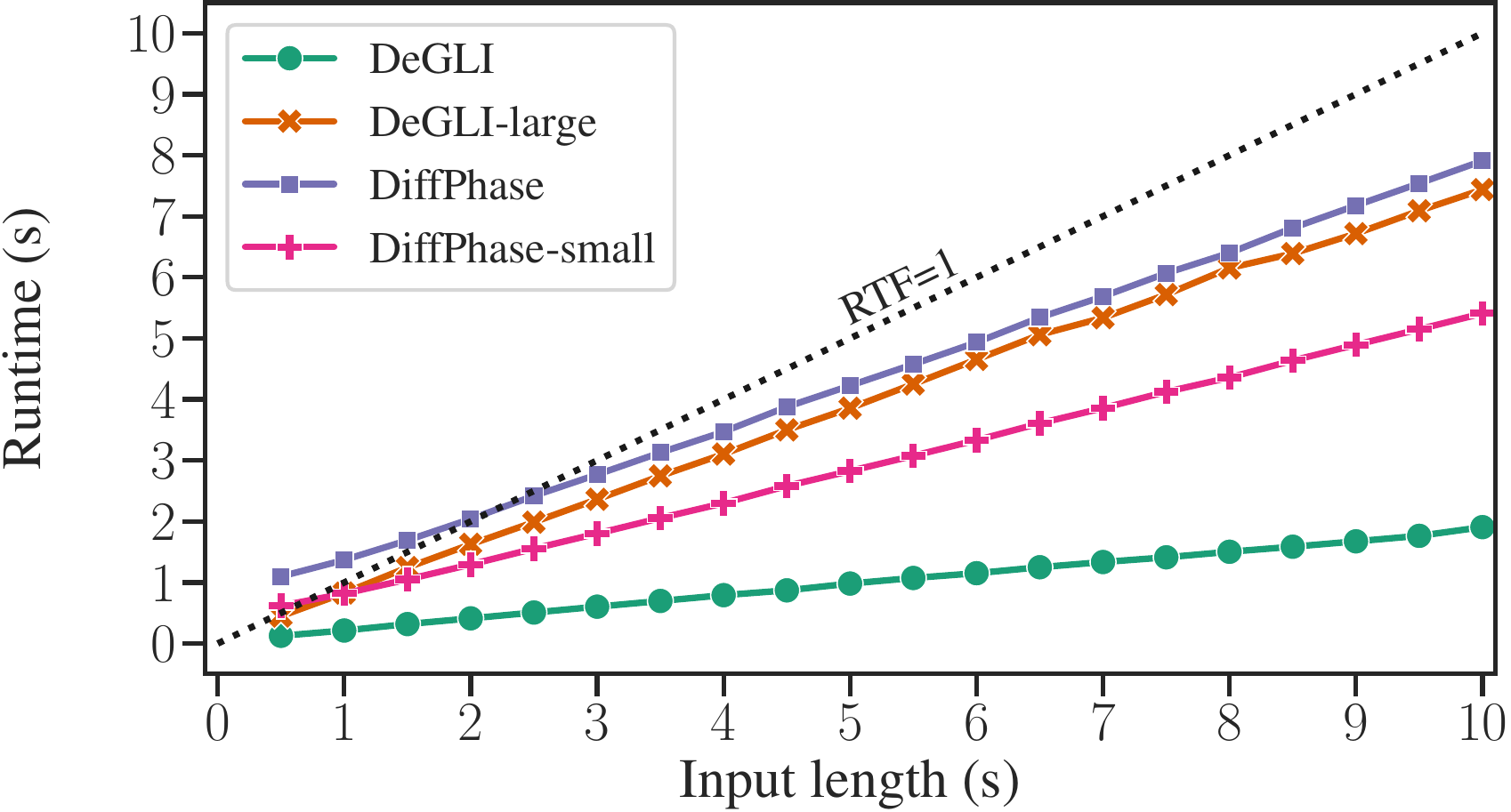}
    \caption{Mean runtime of different methods as a function of the input length. The dotted line represents a real-time factor (RTF) of 1.}\label{fig:runtime}
\end{figure}

A smaller score estimation network (DiffPhase-small) yields a slightly degraded reconstruction but still results in very good performance, superior to DeGLI and GLA. Enlarging the DeGLI network does indeed improve its performance, however only slightly. Unlike DiffPhase, whose performance saturates quickly for larger $N$, the PESQ and ESTOI scores achieved by DeGLI appear to monotonically increase with $N$.  However, for both DiffPhase and DeGLI, the choice of $N$ is directly related to the number of DNN evaluations and therefore affects the overall runtime of each algorithm, which grows linearly in $N$. In \cref{fig:runtime} we compare the measured runtime of the different algorithms for different input lengths and $N=30$, averaged over 100 runs for each algorithm. Although the larger DeGLI model only has about one-tenth as many parameters as the smaller DiffPhase model, it is significantly slower and in fact only slightly faster than the larger DiffPhase model. Considering the performance gap observed at $N=30$ in \cref{fig:pesq_estoi}, we infer that the DiffPhase models achieve superior phase retrieval quality at similar or lower runtime, compared to DeGLI.
\section{Conclusion}
\label{sec:conclusion}
Inspired by the recent success of diffusion models in various fields, we propose to employ a diffusion model for STFT phase retrieval. By considering the loss of phase information as a corruption process, we integrate it into the existing forward stochastic process formulation given by the OUVE SDE. The corresponding reverse process is then solved with the help of a DNN and a numerical SDE solver, yielding an estimate of the unknown phase spectrogram. Experiments on phase reconstruction of clean speech samples show excellent performance in terms of instrumental speech quality and intelligibility metrics, surpassing the performance of another DNN-based method with a comparable computational load.
\clearpage
\section{References}\vspace{-5px}
\label{sec:refs}
\atColsBreak{\vskip1pt}
\printbibliography[heading=none]
\end{document}